\documentclass[12pt,thmsa]{article}
\usepackage{sw20lart}

\input{tcilatex}
\begin{document}

\author{Emilio Santos \and Departamento de F\'{i}sica. Universidad de Cantabria.
Santander. Spain}
\title{Metric fluctuations, solution to the superluminal neutrino problem?}
\date{January, 7, 2012 }
\maketitle

\begin{abstract}
It is shown that the measured neutrino velocity, apparently violating
relativity theory, is compatible with (general) relativistic causality
provided that we assume that the metric of spacetime fluctuates at short
distances.

PACS: 13.15.+g; 04.60.-m; 04.20.Cv
\end{abstract}

The recent measurement\cite{Opera} of a neutrino velocity, $v$, greater than
the velocity of light, $c$, by 
\begin{equation}
(v-c)/c=(2.37\pm 0.32(stat.)_{-0.24}^{+0.34}(sys.))\times 10^{-5},  \label{0}
\end{equation}
has been interpreted as a violation of (special) relativity theory. The
purpose of this note is to show that the result does not imply a violation
of general relativity. In fact if we assume that the metric of spacetime is
fluctuating the result may be explained within that theory.

Let us assume that the neutrinos travel in a spacetime region with a metric 
\begin{equation}
ds^{2}=g_{\mu \nu }dx^{\mu }dx^{\nu },  \label{1}
\end{equation}
where I consider a signature (+ - - -). In order to preserve relativistic
causality we must demand that every particle travels along a timelike path,
which means 
\begin{equation}
ds^{2}>0,  \label{2}
\end{equation}
at all times.

The main assumption in this note is that the metric eq.$\left( \ref{1}%
\right) $ is fluctuating. We might consider quantum fluctuations, which
would mean that the coefficients $g_{\mu \nu }$ of the metric are operators
whose vacuum expectation values have the property 
\begin{equation}
\left\langle vac\left| \left( g_{\mu \nu }\right) ^{2}\right|
vac\right\rangle >\left\langle vac\left| g_{\mu \nu }\right|
vac\right\rangle ^{2},  \label{3}
\end{equation}
for at least one of the metric coefficients. An ``equal'' sign, rather than
``greater'', for all metric coefficients in eq.$\left( \ref{3}\right) ,$
would corresponding to a non-fluctuating metric. Of course the existence of
quantum vacuum fluctuations of all kinds of fields including gravity is not
strange, it belongs to the current paradigm of quantum theory.

We might also study classical fluctuations (due, for instance, to Earth
density fluctuations along the neutrino paths). In this case we should
consider a set of metrics with a probability distribution defined on the
set. In practice the analysis of the neutrino experiment is made as if the
travel takes place in an effective (non-fluctuating) metric which, with good
approximation, may be taken as Minkowskian. Thus the effective metric should
correspond to the average of the true metric, eq.$\left( \ref{1}\right) ,$
that is 
\begin{equation}
\left\langle g_{00}\right\rangle =1,\left\langle g_{0k}\right\rangle
=0,\left\langle g_{ik}\right\rangle =-\delta _{ik},  \label{4}
\end{equation}
where $i,k=1,2,3$ and $\delta _{ik}=1$ if $i=k$, zero otherwise. Here $%
\left\langle g\right\rangle $ may be interpreted as a vacuum expectation (if
we consider quantum fluctuations) or as an ensemble average (if we consider
classical fluctuations).

With an appropriate choice of spacetime coordinates I may assume that a
neutrino travels from the point $x=y=z=0$ at time $t=0$ to the point $%
x=L,y=z=0$ at time $t=T$, with an obvious notation. The relevant point is
that it is possible to have $L>cT$ without violating the inequality $\left( 
\ref{2}\right) $. I will prove this claim presenting a simple example,
admittedly contrived. In order to avoid difficulties with the quantization
of general relativity the model will correspond to classical fluctuations.

In the model I assume that only $g_{00}\left( t\right) $ and $g_{11}\left(
t\right) $ fluctuate, with the remaining $g_{\mu \nu }$ as in Minkowski
metric. In addition I assume that $g_{00}$ and $g_{11}$ depend only on time,
but not on the spatial coordinates. Then the motion of a neutrino may be
derived from the condition of stationary action, that is 
\begin{equation}
\int ds=\int_{0}^{T}\sqrt{g_{00}\left( t\right) -\left| g_{11}\left(
t\right) \right| \stackrel{\cdot }{x}^{2}-\stackrel{\cdot }{y}^{2}-\stackrel{%
\cdot }{z}^{2}}dt=extremum.  \label{5}
\end{equation}
The Euler-Lagrange equations of this variational problem are 
\begin{equation}
\frac{d}{dt}\left( \frac{g_{11}\left( t\right) \stackrel{\cdot }{x}^{2}}{%
\stackrel{\cdot }{s}}\right) =\frac{d}{dt}\left( \frac{\stackrel{\cdot }{y}%
^{2}}{\stackrel{\cdot }{s}}\right) =\frac{d}{dt}\left( \frac{\stackrel{\cdot 
}{z}^{2}}{\stackrel{\cdot }{s}}\right) =0,  \label{6}
\end{equation}
where $\stackrel{\cdot }{s}$ stads for the square root present in eq.$\left( 
\ref{5}\right) .$ The solution is 
\begin{equation}
y=z=0,\stackrel{\cdot }{x}=\sqrt{\frac{g_{00}\left( t\right) }{\left|
g_{11}\left( t\right) \right| +K\left| g_{11}\left( t\right) \right| ^{2}}},
\label{7}
\end{equation}
where the boundary conditions above stated have been taken into account in
the former two equalities. $K>0$ is an integration constant which would
approach $zero$ in the ultrarrelativistic regime. It is easy to see that the
solution eq.$\left( \ref{7}\right) $ guarantees the inequality $\left( \ref
{2}\right) ,$ that is the motion is along a timelike path at all times. The
question which remains is whether the solution is compatible with all
conditions above stated, in particular 
\begin{equation}
\int_{0}^{T}\stackrel{\cdot }{x}dt=L>cT,\int_{0}^{T}\left| g_{11}\left(
t\right) \right| dt=T,\int_{0}^{T}g_{00}\left( t\right) dt=T.  \label{8}
\end{equation}
Actually the latter two conditions correspond to assuming that time averages
may be substituted for the ensemble averages eqs.$\left( \ref{4}\right) ,$
which is plausible.

The compatibility between eqs.$\left( \ref{7}\right) $ and $\left( \ref{8}%
\right) $ may be proved by choosing 
\begin{equation}
\left| g_{11}\left( t\right) \right| =1+A\cos \left( \omega t\right) ,%
\stackrel{\cdot }{x}=\frac{L}{T}+B\cos \left( \omega t\right) ,  \label{9}
\end{equation}
$A$ and $B$ being constant smaller than unity. The term with $\cos \left(
\omega t\right) $ simulates the metric fluctuations. Hence, taking eqs.$%
\left( \ref{7}\right) $ into account, I get 
\begin{eqnarray}
g_{00}\left( t\right)  &=&\left[ 1+K+AK\cos \left( \omega t\right) \right]
\left[ 1+A\cos \left( \omega t\right) \right] \left[ \frac{L}{T}+B\cos
\left( \omega t\right) \right] ^{2}.  \nonumber \\
&=&\left[ 1+K+\frac{1}{2}AK^{2}+A(1+K)\cos \left( \omega t\right) +\frac{1}{2%
}AK^{2}\cos \left( 2\omega t\right) \right] \times   \nonumber \\
&&\times \left[ \frac{L^{2}}{T^{2}}+\frac{1}{2}B^{2}-\frac{2BL}{T}\cos
\left( \omega t\right) +\frac{1}{2}B^{2}\cos \left( 2\omega t\right) \right]
,  \label{10}
\end{eqnarray}
where I use units such that $c=1$. It is easy to see that eqs.$\left( \ref{9}%
\right) $ imply the first two eqs.$\left( \ref{8}\right) $ if we take into
account that we may approximate 
\begin{equation}
I\equiv \int_{0}^{T}\cos \left( \omega t\right) dt\simeq 0,  \label{11}
\end{equation}
which is plausible if $\omega ^{-1}<<T$. Proving the latter eq.$\left( \ref
{8}\right) $ is straightforward if we assume, in addition to eq.$\left( \ref
{11}\right) ,$%
\begin{eqnarray}
\int_{0}^{T}\cos \left( 2\omega t\right) dt &\simeq &0,\int_{0}^{T}\cos
\left( 4\omega t\right) dt\simeq 0,  \nonumber \\
\int_{0}^{T}\cos \left( \omega t\right) dt\cos \left( 2\omega t\right) dt
&\simeq &0  \label{12}
\end{eqnarray}
After some algebra we get from eq.$\left( \ref{10}\right) $%
\begin{eqnarray*}
T^{-1}\int_{0}^{T}g_{00}\left( t\right) dt &=&\left( 1+K+\frac{1}{2}%
AK^{2}\right) \left( \frac{L^{2}}{T^{2}}+\frac{1}{2}B^{2}\right)  \\
&&-A(1+K)\frac{BL}{T}+\frac{1}{8}AK^{2}B^{2}
\end{eqnarray*}
where eqs.$\left( \ref{11}\right) $ and $\left( \ref{12}\right) $ have been
taken into account. In fact it is a simple matter to get values of the
parameters $A,B$ and $K$ able to reproduce both the latter eq.$\left( \ref{8}%
\right) $ and the measured neutrino velocity eq.$\left( \ref{0}\right) $
which corresponds to $L/T\simeq 1+2.4\times 10^{-5}.$ This is the case for
instance if we choose 
\[
K<<A\simeq B\simeq 0.01.
\]

In conclusion I claim that the apparent superluminal velocity of the
neutrinos, recently measured, is compatible with relativistic causality.

\end{document}